\documentclass[aps,prl,twocolumn,amssymb, showpacs,groupedaddress]{revtex4}
\usepackage{graphicx}
\usepackage{dcolumn}

\begin{document}


%
%

\title{The unusual thickness dependence of
superconductivity in $\alpha$-MoGe thin films}

%
%

\author{H. Tashiro$^1$}
\author{J. M. Graybeal$^1$}
\author{D. B. Tanner$^1$}
\author{E. J. Nicol$^2$}
\author{J. P. Carbotte$^3$}
\author{G. L. Carr$^4$}
\affiliation{$^1$Department of Physics, University of Florida, Gainesville, Florida 32611}
\affiliation{$^2$Department of Physics, University of Guelph,
Guelph, Ontario N1G 2W1 Canada}
\affiliation{$^3$Department of Physics and Astronomy, McMaster
University, Hamilton, Ontario N1G 2W1 Canada}
\affiliation{$^4$Brookhaven National Laboratory, Upton, New York 11973}

\date{\today}


%
%

\begin{abstract}
Thin films of $\alpha$-MoGe show progressively reduced $T_{c}$'s as
the thickness is decreased below 30 nm and the sheet resistance
exceeds 100 $\Omega/\Box$. We have performed far-infrared
transmission and reflection measurements for a set of $\alpha$-MoGe
films to characterize this weakened superconducting state. Our
results show the presence of an energy gap 
with ratio $2\Delta_0/k_BT_{c} = 3.8 \pm 0.1$ in all films studied,  
slightly higher
than the BCS value, even
though the transition temperatures decrease significantly as film
thickness is reduced. The material properties follow BCS-Eliashberg theory
with a large residual scattering rate except that the coherence
peak seen in the optical scattering rate is found to be strongly smeared out
in the thinner superconducting samples. A peak in the
optical mass renormalization at $2\Delta_0$ is predicted and 
observed for the first time.
\end{abstract}

%
%


\maketitle

%
%

Disorder and reduced dimensionality
affect the physical properties of metallic systems in a
number of ways. Anomalous diffusion leads to localization of
electrons and a related enhancement of the Coulomb interaction via
reduced screening\cite{anderson79,altshuler80}, seen as an 
increase in $\mu^{*}$, the
renormalized Coulomb interaction parameter. In a system of
lower dimensions, the coupling to disorder increases, and pronounced
effects are expected. Disorder-driven localization and the related
enhancement of the Coulomb interaction inherently compete with the
attractive interaction in superconducting metals\cite{maekawa81,
anderson83}, described by the electron-phonon
spectral density $\alpha^2F(\omega)$\cite{carbotte}. 
This competition reduces the transition temperature. Of
particular interest are two-dimensional (2D) superconductors in
which the degree of disorder can be adjusted by varying the
appropriate parameters. In an ideal 2D system, the relevant
parameter is normally considered to be the sheet resistance,
$R_{\Box}$.  The sheet resistance is determined by two factors: the
(possibly thickness dependent) conductivity $\sigma$ and the film
thickness $d$.

Amorphous MoGe ($\alpha$-MoGe) thin films are thought to be a model
system for studying the interplay between superconductivity and
disorder. Several transport experiments have revealed a sharp
reduction in the superconducting transition temperature $T_{c}$ with
increasing $R_{\Box}$, even in the weakly localized
regime\cite{graybeal84, graybeal85,   strongin,
raffy}. The suppression of $T_c$ has been attributed to
localization and an increase in the Coulomb
interaction\cite{maekawa81}. In this Letter, we explore the $T_c$
suppression in $\alpha$-MoGe thin films with different thickness via temperature-dependent
far-infrared transmittance and reflectance. A strong suppression of $T_{c}$
with increasing $R_{\Box}$ is observed. The superconducting energy gap is
also reduced, but the ratio of gap energy to transition temperature
and the normal-state conductivity, both of which could be dependent on
the disorder-driven Coulomb interaction, are not affected at all.


%

Our films were prepared by co-magnetron sputtering from elemental
targets onto rapidly rotating (3 rev/sec or 1\AA\ deposited/rev)
single-crystal r-cut sapphire substrates (1 mm thick). A 75\AA\
$\alpha$-Ge underlayer was first laid down on the substrates to
ensure smoothness of the subsequently deposited MoGe films. 
For films prepared
in similar fashion, no sign of crystalline inclusions were observed
by x-ray and transmission electron microscopy. This procedure is
known to yield uniform and homogeneous amorphous films of near ideal
stoichiometry\cite{graybeal84, graybeal85}. A
thickness monitor gave the film thickness; the remaining  parameters
of our films, in Table~\ref{tab:table1}, were all determined from
optical measurements, described below.
\begin{table}
\caption{\label{tab:table1} Parameters for MoGe films. }
\begin{ruledtabular}
\begin{tabular}{ccccccc}
Film & $d$ (nm) & $T_{c}$ (K) & $R_{\Box}$ $(\Omega)$ & $2\Delta_{0}$ (cm$^{-1})$ & 
$2\Delta_{0}/kT_{c}$&$n_{s}\footnotemark[1]$\\
\hline
A & 4.3 & $<1.8$ & 505 & - & -& -\\
B & 8.3 & 4.5 & 260 & 12 & 3.7& 1.20\\
C & 16.5 & 6.1 & 131 & 16 & 3.9& 1.49\\
D & 33 & 6.9 & 69 & 18 & 3.8& 1.66\\
\end{tabular}
\end{ruledtabular}
\footnotetext[1]{In units of $10^{21}$/cm$^3$}
\end{table}

Far-infrared measurements were performed at beamlines U10A and U12IR
of the National Synchrotron Light Source at Brookhaven National
Laboratory. U12IR, equipped with a Sciencetech SPS200
Martin-Puplett interferometer, was used for frequencies between
5 and 50 cm$^{-1}$. A Bruker IFS-66v/S rapid scan Fourier-transform
interferometer at U10A was used over 20--100 cm$^{-1}$. A bolometer
operating at 1.7 K provided excellent sensitivity; its window is
responsible for the high-frequency cutoff of 100 cm$^{-1}$. The
films were in an Oxford Instruments Optistat bath cryostat, which
enabled sample temperatures of 1.7--20~K. Transmittance $T(\omega)$
and reflectance $R(\omega)$ of four films were taken at various
temperatures below $T_{c}$. The normal-state transmittance and
reflectance were taken at 10 K.


%
%

For a metal film of thickness $d \ll \lambda$, the wavelength of the
far-infrared radiation, and $d \ll \{\delta, \lambda_{L}\}$, the
skin depth (normal state) or penetration depth (superconducting
state), the transmittance across the film into the substrate and the
single-bounce reflectance from the film are both determined by the
film's complex conductivity $\sigma=\sigma_{1}+i\sigma_{2}$ according 
to\cite{palmertinkham, gao96prb} 
\begin{equation}
\label{Eq:FilmTransmittance}
 T_{f} = \frac{4n}
{(Z_{0}\sigma_{1}d +n+1)^{2}+ (Z_{0}\sigma_{2} d)^{2}},
\end{equation}
\begin{equation}
\label{Eq:FilmReflectance}
 R_{f} = \frac{(Z_{0}\sigma_{1}d + n- 1)^{2} + (Z_{0}\sigma_{2}d )^{2}} 
{( Z_{0}\sigma_{1}d +n+1)^{2}+ (Z_{0}\sigma_{2} d)^{2}},
\end{equation}
where $n$ is the refractive index of the substrate
   and
 $Z_{0}$ is the impedance of free space ($4\pi/c$ in cgs; 377
$\Omega$ in mks). Although Eqs.~\ref{Eq:FilmTransmittance} and
\ref{Eq:FilmReflectance} describe the physics of the thin film on a
thick substrate, the  transmittance and reflectance
are influenced by multiple internal reflections within the 
substrate and the (weak) substrate absorption. After accounting for these 
effects\cite{gao96prb}, measurements of $T$ and $R$ at each
frequency determine $\sigma_{1}$ and $\sigma_{2}$. Beginning with
Palmer and Tinkham\cite{palmertinkham}, this
approach has been used a number of times in the past to obtain the
optical properties of superconducting thin films.

We used the broadband far-infrared transmittance to determine the
transition temperature. The normal-state transmission is temperature
independent (on account of the dominant residual scattering).
When, as the sample temperature is decreased slowly,
superconductivity occurs, the broadband transmission increases. We
call $T_c$ that temperature at which a measurable transmission
increase first occurs. Finally, the normal-state infrared
transmission, via Eq.~(1), gives $R_\Box = 1/\sigma d$. (The
frequency-independent transmission tells us that the normal-state 
$\sigma_1(\omega) = \mbox{constant} \gg \sigma_2$.)


%
%

\begin{figure}
\includegraphics[width=0.99\hsize]{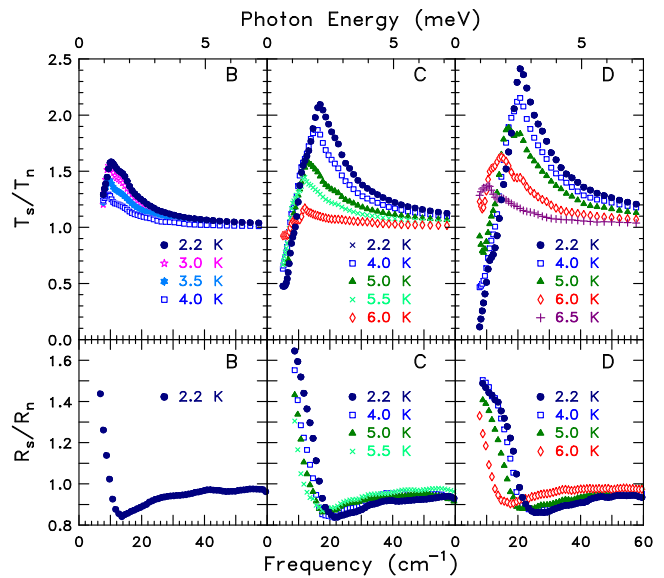}
\caption{\label{fig:ratio} (Color online)
Measured transmittance and reflectance ratios of three 
MoGe films at several temperatures.}
\end{figure}

Figure~\ref{fig:ratio} shows $T_{s}/T_{n}$ and
$R_{s}/R_{n}$ at several temperatures for three films; the thinnest
film did not superconduct at the lowest achievable temperature in
our apparatus.
The shape of the transmission curve is determined by a competition
between $\sigma_{1}$ and $\sigma_{2}$ in
Eq.~(\ref{Eq:FilmTransmittance}). At low frequency the
ratio goes to zero as $\sim\omega^{2}$ due to the kinetic inductance
of the superfluid, which yields $\sigma_{2s} \sim 1/\omega$ while at
the same time $\sigma_{1s} \sim 0$. The frequency of the maximum of
$T_{s}/T_{n}$ occurs very close to the superconducting gap frequency
$\omega_{g}=2\Delta/\hbar$ because $\sigma_{1s}$ rises toward the 
normal-state value above the gap.
At high frequencies $T_{s}/T_{n} = 1$.

The data in Fig.~\ref{fig:ratio} clearly show that the gap shrinks as
temperature increases toward $T_{c}$. At a given reduced temperature
$T/T_{c}$, the gap shifts to lower energy as the film becomes
thinner. The suppression of $T_{c}$ with decreasing thickness 
(increasing $R_{\Box}$) is confirmed as well. Fits to these data using the 
dirty-limit, finite-temperature  Mattis-Bardeen (MB)\cite{mattis}
 conductivity
expressions were good, consistent with the signal-to-noise ratio in the
data, giving 
$2\Delta_{0}/k_{B}T_{c} = 3.8 \pm 0.1$, slightly higher than the BCS
weak coupling limit of 3.5. Changes in $2\Delta_{0}/k_{B}T_{c}$ with
thickness are much smaller than the $T_{c}$ reduction and not
monotonic.  (See Table~\ref{tab:table1} for the fit results.)

\begin{figure}
\includegraphics[width=0.7\hsize]{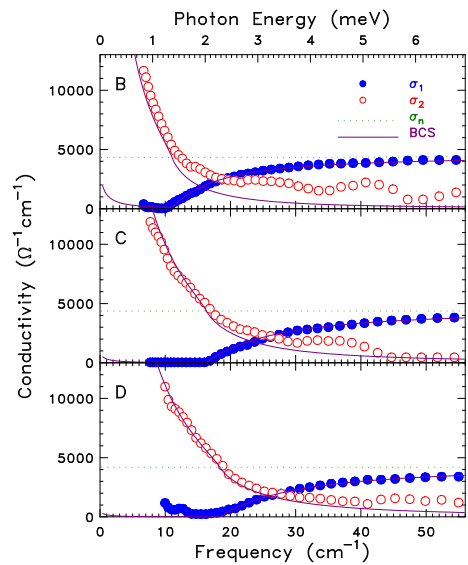}
\caption{\label{fig:conductivity} (Color online)
Real (filled circles) and imaginary (open circles) parts of the
optical conductivity for three $\alpha$-MoGe thin films. The data
are taken at 2.2 K. The Mattis-Bardeen conductivity is also shown.}
\end{figure}

Figure~\ref{fig:conductivity} shows the 2.2 K results for the real
and imaginary parts of the optical conductivity,
$\sigma_{1}(\omega)$ and $\sigma_{2}(\omega)$, for each film.
The Mattis-Bardeen conductivities are also shown. 
 The gap of $2\Delta$ in the absorption spectrum is 
evident.  All three films have
approximately the same normal state conductivity, $\sigma_{N} \sim
4000$~$\Omega^{-1}$cm$^{-1}$ obtained from transmittance measurement
of film in the normal state; the superconducting-state
$\sigma_1(\omega)$ approaches this value at high frequencies. A
similar value (4080~$\Omega^{-1}$cm$^{-1}$) is found by transport
measurements. Thus, we conclude that the normal-state conductivity
(or resistivity) is independent of the thickness of the film.

\begin{figure}
\includegraphics[width=0.7\hsize]{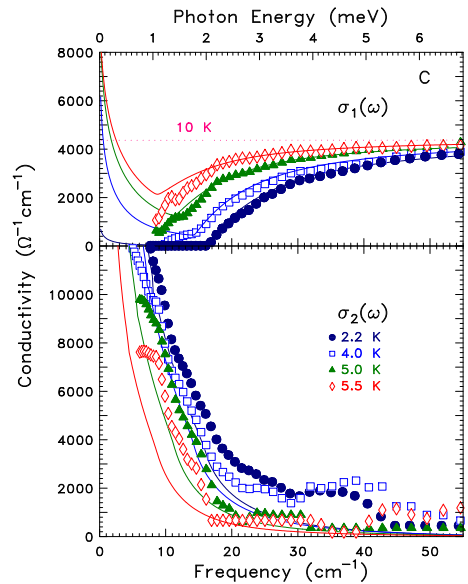}
\caption{\label{fig:conductivity165} (Color online) 
$\sigma_{1}(\omega)$ and $\sigma_{2}(\omega)$ at various
temperatures for the 16.5 nm MoGe film. The points are the
data and the lines are the results of our Eliashberg calculations.}
\end{figure}

As the data is clearly in the dirty limit, the fitting with the MB expressions
is quite adequate for obtaining the value of $\Delta_0$. However, in
order to elucidate further features of the data, discuss changes in
$T_c$, and make predictions, we will now move to more sophisticated
calculations using BCS-Eliashberg theory.
Figure~\ref{fig:conductivity165} shows the results for the real
and imaginary parts of the optical conductivity,
$\sigma_{1}(\omega)$ and $\sigma_{2}(\omega)$, for film C.
The lines are results of numerical calculations for the conductivity
based on the Eliashberg equations and the Kubo formula for the current-current
correlation function\cite{marsiglio}. 
The electron-phonon spectral function was taken
from that obtained through inversion of tunneling data on amorphous Mo\cite{kimli}
and its mass enhancement parameter $\lambda$ is fixed at 0.9. The Coulomb
repulsion $\mu^*$ was adjusted to obtain the measured value of $T_c$.
Other parameters are the impurity scattering rate $1/\tau^{imp}=3.5$ eV
and the plasma energy $\Omega_p=10.7$ eV. We will see later how these were
obtained from the conductivity data itself. The agreement with the data
for $\sigma_1$ is best at the lowest temperature considered, with small
deviations for $T$ near $T_c$. This is true for all three films.
The theory for  $\sigma_{2}$ agrees with the $\sim
1/\omega$ low-frequency behavior but tends to be below the
experiment, especially at higher frequencies. The fit is less good with
increasing $T$, although the qualitative behavior is given 
correctly.

\begin{figure}
\includegraphics[width=0.6\hsize]{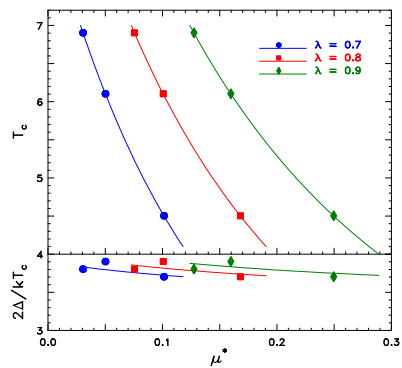}
\caption{\label{fig:tcmustar} (Color online)
Dependence of $T_c$ and $2\Delta/k_BT_c$ on Coulomb
repulsion $\mu^*$ for three values of electron-phonon mass enhancement 
$\lambda$.}
\end{figure}

As changing the thickness of the sample could change both $\mu^*$ and the
electron-phonon interaction,
there is some choice in fitting the data with Eliashberg theory. 
In Fig.~\ref{fig:tcmustar}, we show results for $T_c$ and the gap ratio
as a function of $\mu^*$ for three values of $\lambda$. 
For fixed $\lambda$, the points on the $T_c$ curve are from the experimental
data for the MoGe films, 
illustrating the $\mu^*$ needed to obtain the $T_c$. With
$\mu^*$ and $\lambda$ now fixed, the experimental points for the gap ratio
can be compared to the prediction and there is good
agreement.
It is clear
from this figure that keeping the ratio at 3.8 can be achieved through 
a change in $\mu^*$ as suggested in\cite{anderson79,
altshuler80,maekawa81,anderson83} but one cannot rule out additional small
changes in $\lambda$. In fact, H\"ohn and Mitrovi\'c\cite{hohn} in their
Eliashberg analysis of tunneling data on disordered Pb films found 
evidence for a change in both these parameters with changing $E_F\tau^{imp}$, 
where
$E_F$ is the Fermi energy. Here such differences will not matter as we
are in an impurity-dominated regime and the optics is not sensitive to
the $\mu^*$ or $\lambda$ value as we will explain. A $\lambda$ of order 1
is needed, however, to get the measured value of the gap to $T_c$.
For definiteness, we only change $\mu^*$ leaving $\alpha^2F(\omega)$ fixed.

\begin{figure}
\includegraphics[width=0.6\hsize]{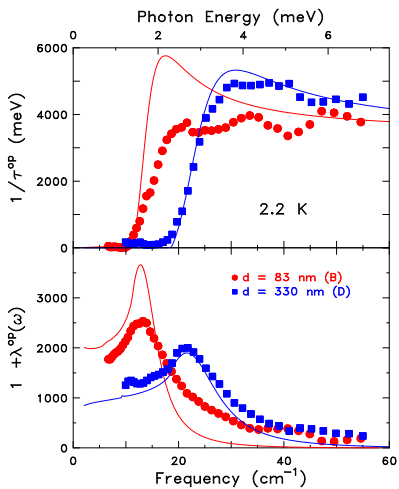}
\caption{\label{fig:tauop}
(Color online)
Optical scattering rate $1/\tau^{op}(\omega)$ and mass renormalization
$1+\lambda^{op}(\omega)$ for the thickest and thinnest superconducting
films. Points are
data and lines are Eliashberg calculations for the extreme dirty 
limit.}
\end{figure}

To proceed with the analysis, we introduce the optical self-energy 
$\Sigma^{op}(T,\omega)$ and use the extended Drude model, where the conductivity
is written 
$\sigma(T,\omega)= (i\Omega_p^2/4\pi)/(\omega-2\Sigma^{op}(T,\omega))$.
The real part of $\Sigma^{op}$ gives the optical mass renormalization
$\lambda^{op}(T,\omega)$ with $\omega \lambda^{op}(T,\omega)=-2
\Sigma_1^{op}(T,\omega)$ and its imaginary part is related to the optical 
scattering rate according to  $1/\tau^{op}(T,\omega)=-2
\Sigma_2^{op}(T,\omega)$.
These quantities are shown in Fig.~\ref{fig:tauop} for the thickest and
thinnest superconducting samples at $T=2.2K$. 
To obtain $1/\tau^{op}$, we had to
use an impurity scattering rate of 3.5 eV.
For $v_F\sim 1.5\times 10^8$ cm/sec, this rate corresponds to a mean free path of
$\sim 0.3$ nm. This value, while small, is consistent with other estimates and is
much less than the thickness of the
films\cite{graybeal84}. Hence surface scattering is not important,
$R_{\Box} \propto 1/d$, and  
the normal-state conductivity does not
depend on sheet resistance. 

It is important to understand that the peaks in $1/ 
\tau^{op}(\omega)$ are
the optical equivalent of density of states coherence peaks. The 
calculation for the
thickest film fits the data well but for the thinner film the peak is
very much attenuated, perhaps indicating a new effect outside standard
Eliashberg theory. In their tunneling study of the metal-insulator
transition in aluminum films, Dynes and coworkers\cite{dynes} found a
similar effect, namely, a broadening of the density of states coherence
peak with increased sheet resistance. The lower panel of 
Fig.~\ref{fig:tauop} gives the optical effective mass in the superconducting
state. For both samples as $\omega\to 0$, this quantity is very large,
of the order of 1000, which is comparable to heavy fermion masses,
although its origin is quite different. 
 These values reflect directly the large impurity
scattering and are related to the decrease in superfluid density with
decreasing $\tau^{imp}$. In an Eliashberg superconductor, the 
superfluid density ($n_s$) at $T=0$ in the clean-limit case is
given by $n_s^{\rm clean}(T=0)=n/(1+\lambda)$, where $n$ is the electron
density. In the dirty limit where $1/[2\Delta_0\tau^{imp}(1+\lambda)]\gg 1$,
it is instead given by $n_s^{\rm dirty}(T=0)=n\pi\Delta_0\tau^{imp}$
with $\lambda$ dropping out\cite{marsiglio}. 
For fixed $n$, this gives immediately a relation between the 
superfluid density $n_s$, $T_c$, and $\sigma_n$ \cite{uemura, homes, argument}
The superfluid density so estimated is shown in Table~\ref{tab:table1};
its variation  is due entirely to the change in $T_c$.
 We note 
also the large peak at $2\Delta_0$ in $1+\lambda^{op}$,
predicted by theory and seen in the 
 data.

In summary, 
the observed strong suppression of $T_c$ with increasing $R_\Box$ while the
ratio $2\Delta_0/kT_c$ remains constant at the intermediate value of 3.8
can be easily accounted for by a small decrease in $\lambda$, an increase
in $\mu^*$ or a combination of both. The 
large residual scattering rate of our  MoGe films makes their optical response  indistinguishable
from BCS; yet, because of Anderson's theorem, $2\Delta_0/kT_c$ remains
bigger than 3.54. In such dirty samples a large value of the optical
effective mass is predicted as well as a peak at $\omega=2\Delta_0$,
with a rapid decrease as $\omega$ is increased. Both effects are observed, the
peak for the first time. 
Moreover, the optical scattering
rate shows the expected coherence peak in the thickest film considered
but is strongly suppressed in the thinnest superconducting 
one. This effect cannot
be understood within BCS-Eliashberg theory and may indicate new physics.


%
%

Research is supported by the U.S. Department
of Energy through Contract DE-FG02-02ER45984 at the University of
Florida, DE-AC02-98CH10886 at the Brookhaven National Laboratory,
NSERC of Canada and the Canadian Institute
for Advanced Research.




\begin{thebibliography}{99}
\bibitem{argument} The ``missing area'' in $\sigma_1(\omega)$ gives  
the weight of the delta function and, consequently, the superfluid density: 
$
4\pi n_se^2/c^2 = 1/\lambda_L^2(T=0)
=\sigma_n 4\pi^2\Delta_0/c^2
= 75 \sigma_n T_c/c^2
$. If $\sigma_n$ is constant, then $n_s \propto T_c$.
\end{thebibliography}
\end{document}